\documentclass[conference]{IEEEtran}
\IEEEoverridecommandlockouts

\usepackage{cite}
\usepackage{float}
\usepackage{stfloats}
\usepackage{booktabs}
\usepackage{multirow}
\usepackage{tabularx}
\usepackage{makecell}
\usepackage{multicol}
\usepackage{bm}
\usepackage{amsmath,amssymb,amsfonts}
\usepackage{algorithmic}
\usepackage{graphicx}
\usepackage{textcomp}
\usepackage{xcolor}

\usepackage{tikz}
\usetikzlibrary{positioning, shapes, arrows.meta}

\def\BibTeX{{\rm B\kern-.05em{\sc i\kern-.025em b}\kern-.08em
    T\kern-.1667em\lower.7ex\hbox{E}\kern-.125emX}}

\usepackage{fancyhdr}
\newcommand{\IEEEcopyrightcover}{%
  \thispagestyle{empty}%
  \onecolumn
  \begin{flushleft}
    \vspace*{0.25cm}
    {\Huge IEEE Copyright Notice}\\[1cm]
    {\normalsize
    © 2026 IEEE. Personal use of this material is permitted. Permission from IEEE must be obtained for all other uses, \\
    in any current or future media, including reprinting/republishing this material for advertising or promotional purposes, \\
    creating new collective works, for resale or redistribution to servers or lists, or reuse of any copyrighted component of \\
    this work in other works.\\[14pt]

    Accepted for publication in the \textit{2026 IEEE International Symposium on Spectrum Innovation (DySPAN 2026)}.\\
    The Version of Record will appear in IEEE Xplore.\\[8pt]

    }
    \vfill
  \end{flushleft}
  \newpage
  \twocolumn
  \setcounter{page}{1} 
}

\begin{document}

\IEEEcopyrightcover

\title{Towards Realistic Waveform-Level IoT Network Simulation via IQ Mixing}

\author{
    \IEEEauthorblockN{Alexis Delplace}
    \IEEEauthorblockA{
        \textit{LISN} \\
        \textit{Université Paris-Saclay}\\
        Gif-sur-Yvette, France \\
        alexis.delplace@lisn.fr
    }
\and
    \IEEEauthorblockN{Samer Lahoud, \IEEEmembership{Senior Member, IEEE}}
    \IEEEauthorblockA{
        \textit{Faculty of Computer Science} \\
        \textit{Dalhousie University}\\
        Halifax, Canada \\
        sml@dal.ca
    }
\and
    \IEEEauthorblockN{Kinda Khawam}
    \IEEEauthorblockA{
        \textit{LISN} \\
        \textit{Université Paris-Saclay}\\
        Gif-sur-Yvette, France \\
        kinda.khawam@lisn.fr
    }

\and
    \IEEEauthorblockN{Dominique Quadri}
    \IEEEauthorblockA{
        \textit{LISN} \\
        \textit{Université Paris-Saclay}\\
        Gif-sur-Yvette, France \\
        dominique.quadri@lisn.fr
    }
}

\maketitle
\thispagestyle{fancy}

\begin{abstract}
Most Internet of Things (IoT) network simulators are packet-level discrete-event systems in which physical-layer (PHY) behavior is approximated through analytical interference rules and precomputed error models. While this enables scalable experiments, it can miss key waveform-level effects such as adjacent-channel leakage, cross-modulation interference between coexisting signals, and receiver imperfections, which are critical in heterogeneous sub-GHz ISM-band coexistence scenarios. This paper discusses these limitations and introduces \emph{IQSim}, a simulation paradigm based on in-phase/quadrature (IQ) stream mixing. Instead of predicting packet outcomes from abstract collision models, IQSim maintains a shared complex baseband \emph{IQStream} into which simulated transmissions are inserted as IQ waveforms after propagation processing, and then demodulated by software-based receivers or hardware gateways. We outline the end-to-end workflow, including online or offline waveform generation, IQ-domain propagation, waveform superposition, and delivery to gateways. We also report preliminary prototype results supporting the feasibility of real-time execution.
\end{abstract}


\vspace{-0.1cm}

\section{Introduction}

The Internet of Things (IoT) has become a cornerstone of modern digital ecosystems, enabling large-scale sensing, monitoring, and control across a wide range of applications. With the rapid growth of intelligent systems, the demand for sensor data is continuously increasing. Data-driven decision-making systems require large volumes of accurate sensor data, necessitating increasingly large and diverse IoT deployments.

To enable these large-scale IoT networks, long-range communication is required. For this reason, low-frequency sub-GHz bands are commonly used, as they provide favorable propagation characteristics over long distances. However, these frequency bands are inherently limited in available bandwidth, and IoT systems typically operate within the unlicensed portions of this spectrum, which are even more constrained. Unlike technologies such as Wi-Fi that expand towards higher frequencies to increase throughput, at the cost of transmission range, IoT technologies cannot easily move upward in the spectrum without sacrificing the propagation properties required for long-range coverage. Regardless of future technological advances, the available spectrum remains a finite resource, making its efficient use a critical challenge.

A broad set of wireless communication technologies tailored to IoT applications already exists \cite{9057670}, and the research community improves them incrementally by proposing small changes aimed at better spectral efficiency and scalability. To validate such proposals and assess their performance, researchers often rely on simulation, because large-scale real-world deployments are prohibitively expensive and, in many cases, not feasible when a mechanism remains a design concept without a production implementation. Beyond this, simulators are also used to estimate the performance of real-world deployments before their actual deployment, and to benchmark existing IoT technologies under various conditions, including their robustness to interference and coexistence with other systems. However, the validity of conclusions drawn from such simulations is only as good as the simulator’s ability to reproduce real-world physical and medium access behaviors.

Most existing IoT network simulators are rooted in packet-level and Medium Access Control (MAC)-centric abstractions, enabling fast and scalable evaluation of network behavior. For studies focused on protocol logic, traffic dynamics, or MAC-layer mechanisms, such abstractions are often sufficient and remain well suited to the problem. However, these simulators do not model actual signal modulation and demodulation, so physical (PHY) layer phenomena such as interference, capture, or synchronization are only approximated. As a result, when conclusions depend on receiver behavior under realistic channel and coexistence conditions, the level of PHY fidelity directly affects the reliability of performance evaluations.

The contribution of this paper is twofold. First, it highlights and discusses the limitations of existing network simulators when used to reproduce real-world IoT system behavior with high physical fidelity, as detailed in Section~\ref{sec:limitations}. Second, it introduces the concept of IQSim, described in Section~\ref{sec:iqsim}, a network simulator based on IQ stream mixing, designed to more realistically reproduce physical layer behavior. This work presents the idea of IQSim as a simulation approach intended to bridge the gap between traditional event-driven models and real-world signal level interactions, enabling more accurate and reliable simulations of IoT systems in complex coexistence scenarios.

\section{Limitations of Existing IoT Network Simulators}
\label{sec:limitations}

\vspace{-0.2cm}

\begin{figure}[H]
    \centering
    \includegraphics[width=0.325\textwidth]{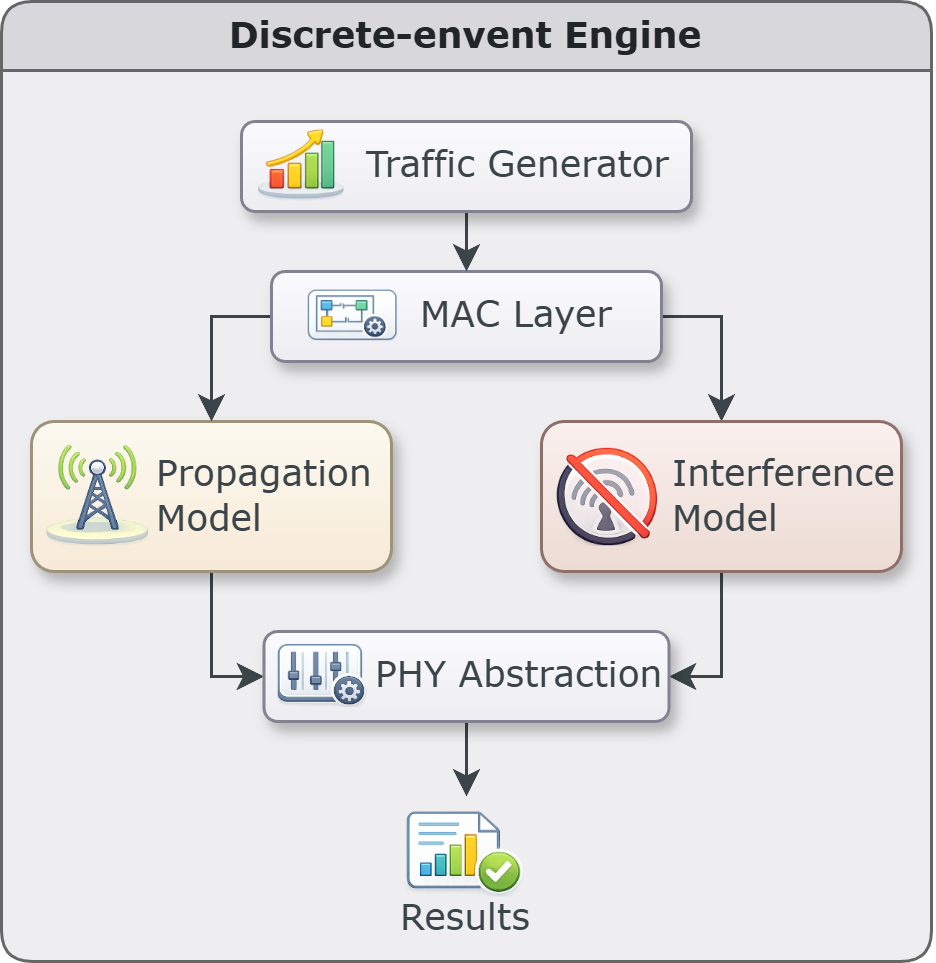}
    \caption{General framework of packet-level wireless network simulators.}
    \label{fig:sim_arch}
\end{figure}

\vspace{-0.1cm}

Most wireless network simulators operate as packet-level discrete-event systems, where the physical layer is abstracted through analytical rules. Regardless of the specific tool, they typically share a common architecture, as illustrated in Fig.~\ref{fig:sim_arch}.

At the core lies a discrete-event engine that orchestrates the simulation timeline and triggers event execution. Traffic generators produce packets following statistical patterns or IoT workloads. The MAC layer governs channel access, scheduling, and retransmission logic. To model the physical environment, a propagation model computes the received signal strength based on path loss, shadowing, and fading, utilizing models such as Okumura-Hata, COST-231, or Log-distance path loss. Simultaneously, an interference model tracks overlapping transmissions to evaluate their combined impact at the receiver. Depending on the abstraction level, this may involve aggregating interfering powers to calculate the Signal-to-Interference-plus-Noise Ratio (SINR) or checking for simple time-frequency overlaps. Finally, the Physical Layer (PHY) abstraction interprets these metrics to approximate demodulation performance. Instead of processing waveforms, it typically maps the SINR to a Packet Error Rate (PER) using pre-computed curves or threshold-based models.

This paradigm is widely adopted because it enables rapid benchmarking and scalable evaluation of collision processes with limited computational overhead. However, when conclusions depend on physical-layer behavior, this efficiency comes at the cost of realism. By approximating radio-frequency interactions rather than emulating them, these simulators rely on idealized assumptions that do not capture the full complexity of the physical layer. Consequently, these abstractions may miss waveform-level effects that directly influence reception outcomes under interference.

\subsection{The Coexistence Blind Spot}

\subsubsection{Intra-Technology Interference}

In standard simulators, intra-technology interference is reduced to collision rules that decide packet reception under overlaps. These rules encode simplified capture, timing, and frequency-selectivity assumptions, which may fail to reflect actual device behavior and lead to inaccurate capacity and reliability estimates. This mismatch becomes apparent when comparing commonly used modeling assumptions with empirical observations from real systems.

For instance, in LoRaWAN, many analytical models assume that Spreading Factors (SF) are strictly orthogonal. Early simulators therefore treated different SFs as non-interfering channels. Experimental studies have shown that SF orthogonality is imperfect due to power disparities and receiver non-idealities~\cite{8267219}. While some simulators mitigate this via empirical collision matrices~\cite{9015882}, many legacy or simplified models still rely on idealized orthogonality.

A similar gap between modeling assumptions and measured behavior exists in sub-GHz Wi-Fi, and specifically in IEEE 802.11ah (Wi-Fi HaLow). The few available PHY-level implementations, such as the ns-3 module in~\cite{tian2018extension}, treat non-overlapping channels as statistically independent under strict spectral-mask compliance, thereby neglecting Adjacent-Channel Interference (ACI). Measurements report substantial mutual interference between adjacent channels, with throughput degradations approaching 50\% when contiguous channels are active in the scenarios evaluated in~\cite{chounos2025scalability}.

Overall, intra-technology interference models based on idealized assumptions tend to underestimate self-interference and fail to capture the abrupt performance degradations observed in practice. While illustrated here with LoRaWAN and IEEE 802.11ah, similar limitations are likely to arise in other sub-GHz IoT technologies when implementation-specific physical-layer effects are not explicitly modeled.

\subsubsection{Cross-Technology Coexistence}

These limitations become more critical in heterogeneous deployments. Most IoT technologies operate in unlicensed ISM bands, where license-free access brings a wide range of technologies such as LoRa, Sigfox, IEEE 802.11ah (Wi-Fi HaLow), IEEE 802.15.4g, Wireless M-Bus, and proprietary protocols into narrow sub-GHz bands. In such environments, cross-technology interference can dominate packet loss and coverage degradation. A simulator that models LoRa collisions accurately but ignores Sigfox or HaLow transmissions will systematically underpredict losses, so realistic evaluation requires explicit cross-technology, not only intra-technology, interference.

Existing tools can be broadly grouped into two categories. On the one hand, custom simulators (e.g., LoRaSim, LoRaEnergySim, MATLAB models) are designed for technology-specific evaluations, but they are intrinsically siloed, implementing a single PHY while treating the rest of the band as static noise or ignoring it altogether. On the other hand, general-purpose discrete-event frameworks (e.g., ns-3, OMNeT++, CupCarbon) provide complete protocol stacks and shared channels in principle, but all modules do not always rely on a unified coexistence interface, which limits realistic cross-technology interactions.

In ns-3, spectral coexistence requires protocol modules to rely on the Spectrum framework, in which transmissions are represented as power spectral densities (PSDs) over a shared channel abstraction, the \texttt{SpectrumChannel} class. However, widely used implementations, such as the SignetLab-DEI LoRaWAN module~\cite{7996384}, rely on custom channel models rather than this class. Its documentation explicitly states that interference from other technologies cannot be accounted for without a unified spectral interface~\cite{signet_lora_doc}. Although Spectrum-based LoRa implementations exist~\cite{reynders2018lorawan}, fragmentation across codebases complicates interoperability and reproducibility.

A similar issue arises in OMNeT++. While the INET framework provides a shared medium abstraction, implemented through the \texttt{RadioMedium} class, for interference calculations, extensions such as FLoRa~\cite{8406255} often implement internal, logic-driven collision rules that do not naturally interact through spectral superposition. Even when a shared medium is used, coexistence is commonly reduced to scalar received-power aggregation and a single interference metric, which ignores signal structure and cannot represent waveform-level effects, such as the partial corruption of a LoRa chirp by a narrowband Sigfox burst~\cite{7500415}.

CupCarbon~\cite{8319179} sits between packet-level simulators and waveform-level tools. It includes explicit modulation and demodulation chains with symbol-level accuracy for technologies such as ZigBee, Wi-Fi, and LoRa. However, interference is modeled as exogenous noise processes, for example AWGN or alpha-stable noise parameterized by node density. Such noise-based abstractions have been shown to produce interference effects that differ from those observed with real coexisting transmissions~\cite{huang2025physical,KavousiGhafi2021}. Consequently, collision outcomes remain driven by probabilistic noise models instead of deterministic interactions arising from the superposition of coexisting waveforms on a shared baseband.

\subsection{The ``Ideal Hardware'' Assumption}

Beyond protocol logic and interference abstractions, most packet-level simulators implicitly assume ideal and homogeneous radio front-ends. In practice, low-cost IoT transceivers exhibit non-negligible device and chipset-dependent effects, including oscillator instabilities such as residual carrier-frequency offset (CFO) and phase noise~\cite{9500616}, limited receiver selectivity and dynamic range, and transmitter nonlinearities that lead to in-band distortion and spectral regrowth, thereby exacerbating adjacent-channel interference. By enabling simulations to be driven by real baseband implementations or real radio hardware, such effects can be naturally reflected in the generated signals, allowing their practical impact on end-to-end performance to be assessed.

\subsection{Limitations of Simulator Extensibility}

Beyond model fidelity and coexistence, keeping IoT simulators aligned with evolving protocols and hardware remains challenging, as it requires continuous integration of specification updates and evolving device characteristics. The slow integration of new PHY features is a concrete example. To the best of our knowledge, no publicly available LoRaWAN network simulator currently implements the full specification including the LR-FHSS modulation introduced in late 2020. While LR-FHSS simulators exist~\cite{10597000}, they remain separate from LoRaWAN network simulators and cannot represent LR-FHSS behavior within realistic deployments.

Rigidity also appears when combining technologies within general-purpose tools. In ns-3, technology stacks not included in the official distribution are developed and released independently, which complicates their joint use within a single simulation. Making two such modules coexist in a single scenario, for example a LoRaWAN implementation and a sub-GHz Wi-Fi model, requires substantial manual effort to manually merge independent modules into a common codebase, aligning build systems, dependencies, and channel abstractions. OMNeT++ and the INET framework adopt a more modular architecture, and technology stacks built on INET can, in principle, be mixed more easily. In practice, cross-technology experiments remain constrained by version-compatibility issues between OMNeT++, INET, and domain-specific extensions.

CupCarbon avoids version conflicts by natively supporting multiple technologies, but users remain restricted to the PHYs implemented by default, and adding a new physical layer still requires substantial development effort due to its explicit modulation and demodulation chains.

Many LPWAN physical layers and some gateway behaviors are proprietary, which limits access to demodulation details and receiver-internal mechanisms. Simulator developers must therefore rely on reverse engineering or sparse empirical measurements, and models often lag behind deployed systems until external studies reveal effects omitted from available documentation, as illustrated by the spreading factor orthogonality assumptions discussed earlier. This also applies to hardware behavior, for example the documentation of a widely used ns-3 LoRaWAN module reports that several aspects of the SX1301 gateway, such as receive path allocation and concurrent packet handling, had to be inferred from incomplete datasheets~\cite{signet_lora_doc}. In contrast, a simulator that can be driven by real devices can evaluate new or closed technologies without waiting for a dedicated, fully validated simulator module.

\section{Concept of IQSim: Toward IQ-Based Wireless Simulation}
\label{sec:iqsim}

A perfectly faithful replication of real-world deployments is unattainable, due to the large number of environmental and hardware-dependent variables. However, the gap between simulation and reality can be reduced by minimizing approximations, in particular those introduced by analytical models of interference and physical-layer behavior. In the approach proposed in this paper, packet outcomes are not predicted through collision rules or precomputed performance curves. Instead, IQSim relies on actual demodulation performed by either hardware gateways or software-based receivers. Because it relies on real receiver implementations, IQSim is best viewed as a wireless \emph{emulator}.

\begin{figure}[H]
    \centering
    \includegraphics[width=0.4\textwidth]{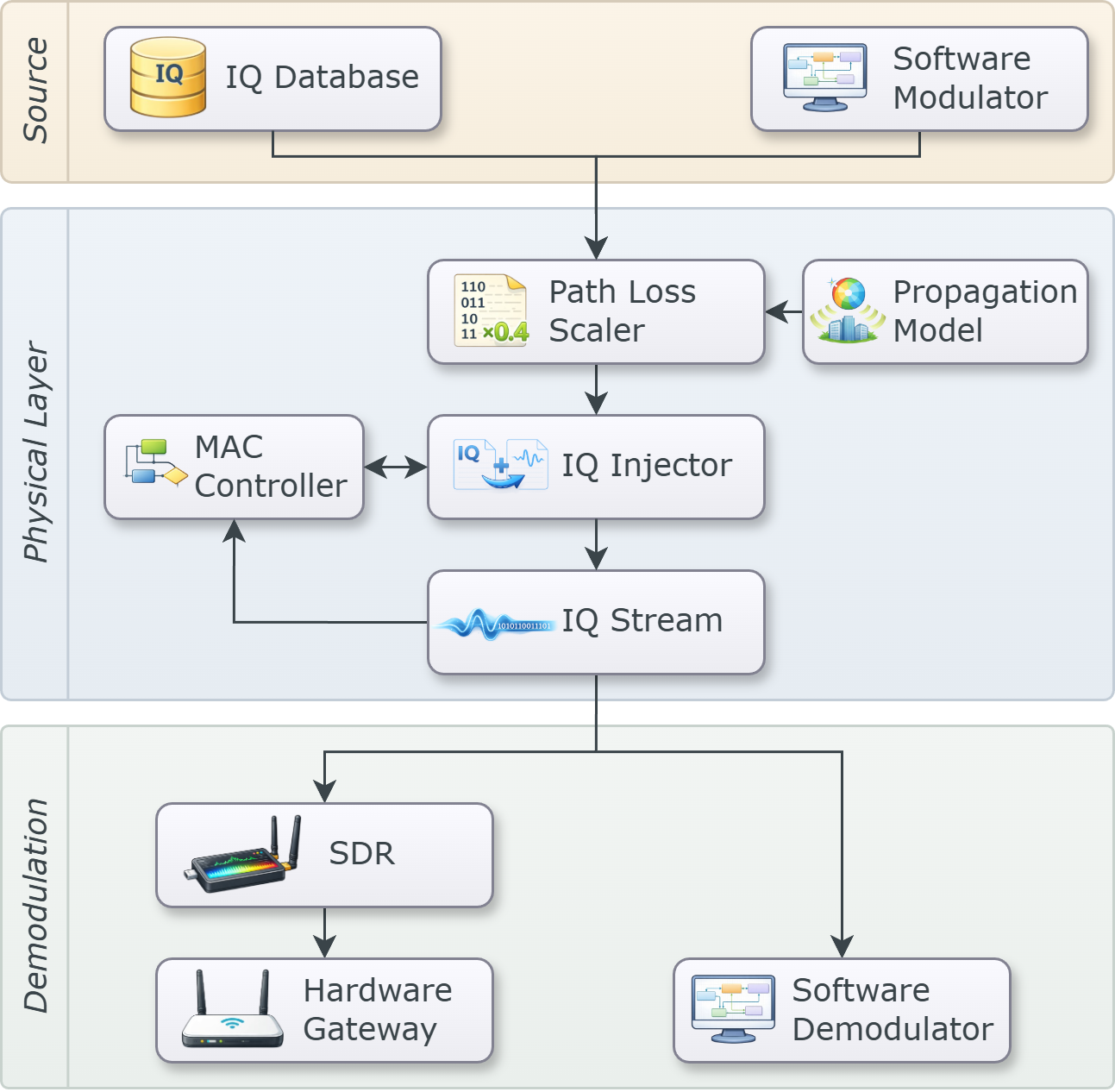}
    \caption{High-level architecture of IQSim illustrating IQ-based signal generation, propagation, mixing, and delivery to gateways.}
    \label{fig:iqsim_schematic}
\end{figure}

The core idea is to maintain a continuous complex baseband stream, referred to as the \emph{IQStream}, representing the whole sub-GHz ISM band or a selected sub-band. Simulated transmissions are generated or replayed as IQ waveforms and inserted into this shared stream. Gateways then process the resulting composite signal as if it were received from the air interface. Figure~\ref{fig:iqsim_schematic} summarizes the end-to-end workflow, from IQ waveform generation and propagation to demodulation.

\subsection{IQ Signal Generation}
\label{subsec:iq_generation}

\vspace{-0.2cm}

In IQSim, IQ waveforms can be produced \emph{online} (in real time during the simulation) or \emph{offline} (generated in advance and replayed during execution). Both modes can rely on either physical devices or software modulators, each with distinct trade-offs.

\emph{Online generation} produces samples as the scenario runs. Its main advantage is seamless integration with realistic applications: payloads can be generated by the same software stack that would produce traffic in an operational deployment and immediately converted into IQ. Two approaches are possible. First, IQ can be generated by \emph{real transmitters} connected to a controlled RF chain. This option is essentially plug-and-play, as it does not require an external software modulator and naturally includes hardware impairments and spectral artifacts. However, this approach directly ties the emulator capacity to the number of available physical devices. Reproducing many concurrent transmissions would require one transmitter per active node, which quickly becomes impractical and defeats the purpose of large-scale simulation.

Second, IQ can be generated using \emph{software-defined modulators}, for instance implemented in GNU Radio. This removes the hard limit imposed by physical hardware and allows a large number of concurrent transmissions to be generated dynamically. Scalability is then constrained by computational resources, since generating many waveforms in real time can become costly depending on bandwidth, sampling rate, and modulation complexity. Moreover, this option requires the existence of an accurate software modulator, which is currently not available for all technologies operating in the sub-GHz ISM band.

\emph{Offline generation} prepares IQ waveforms ahead of time and replays them during simulation. The primary benefit is that waveform generation or capture does not need to satisfy real-time constraints, significantly reducing computational pressure during execution. Offline IQ can be obtained by recording IQ traces using a Software-Defined Radio (SDR) from real transmitters in a controlled RF environment, allowing overlapping transmissions to be reproduced without multiple live radios at runtime. Alternatively, waveforms can be generated offline using software modulators, offering the same modeling capabilities as their online counterparts while shifting the computational cost to a preprocessing phase. The main limitation of offline approaches is the storage requirement associated with large IQ datasets. Each complex sample is stored as an \mbox{(I,Q)} pair of 16-bit integers, \textit{i.e.,} 4~bytes per sample; consequently, file size scales linearly with sampling rate and signal duration. Depending on these parameters, a single IQ trace typically ranges from a few hundred kilobytes to several tens of megabytes. Replay-based methods also remain restricted to waveforms that have been generated or recorded in advance.

\subsection{Propagation and IQStream Injection}
\label{subsec:propagation_injection}

For each individual transmission IQ waveform, the emulator applies a propagation model to emulate distance-dependent attenuation, fading, and other environment-induced effects. A basic implementation computes an attenuation factor from a conventional path-loss model and applies it directly to the IQ samples. More advanced approaches may incorporate ray-tracing engines (\textit{i.e.,} Sionna RT\cite{sionna}) or time-varying scaling functions to reflect mobility. Since propagation is applied in the IQ domain, the resulting distortions are preserved in the waveform delivered to each gateway.

After propagation, the IQ signal is injected into the \emph{IQStream} associated with that gateway. Because the \emph{IQStream} aggregates all simulated devices, interference and collisions emerge naturally from IQ-domain signal superposition. This superposition also aggregates noise components present in individual signals, which highlights the importance of well-controlled signal sources when generating or recording IQ waveforms.

\subsection{End-to-End Delivery to Gateways}
\label{subsec:delivery_gateways}

An \emph{IQStream} is generated for each gateway, only co-located gateways process the same composite waveform.

For \emph{software-based gateways}, the stream can be forwarded as a continuous flow of complex samples. The demodulation chain then operates as if it were fed with live baseband samples.

For \emph{hardware gateways}, the IQStream is sent to an SDR device that performs digital-to-analog conversion and RF reconstruction. The SDR output is fed to the gateway through a coaxial cable, avoiding uncontrolled environmental variability.

In both cases, each gateway demodulates the waveform generated for it, so packet outcomes directly reflect receiver behavior under realistic interference conditions.

\subsection{Prototype and Preliminary Feasibility}
\label{subsec:prototype_feasibility}

To validate that IQSim is practically achievable, we implemented a first prototype and evaluated it on a realistic IoT setup operating over a 1.5~MHz band. Using the same processing pipeline, the prototype sustains real-time execution on a modern CPU with up to 100 concurrent devices. With GPU parallelization, it scales to tens of thousands of devices in our initial experiments while preserving real-time operation, demonstrating practical waveform-accurate IoT emulation on off-the-shelf hardware.

\begin{figure}[H]
  \centering
  \includegraphics[width=\linewidth]{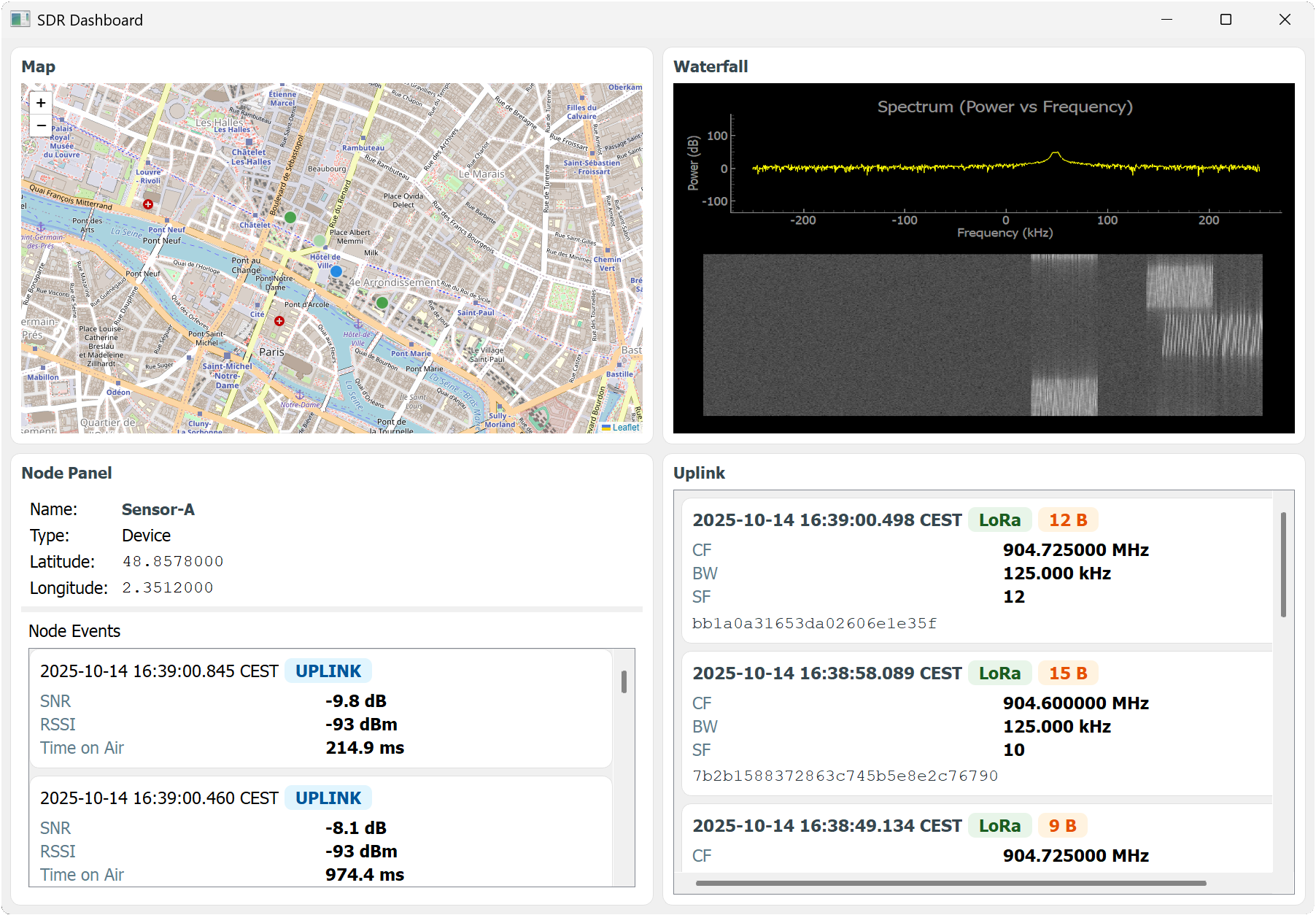}
  \caption{Prototype IQSim used in the preliminary feasibility experiments.}
  \label{fig:iqsim_gui}
\end{figure}

These results depend on the resampling method and sampling rate, since each signal must be resampled to the IQStream rate before injection. While these measurements indicate feasibility rather than a definitive upper bound, they suggest that IQSim is already practical, at least with offline IQ generation. Important next steps are to extend the evaluation to online signal generation, validate the simulator against real-world measurement data, and quantify overhead relative to traditional simulators.

\section{Conclusion}
\label{sec:conclusion}

IQSim suggests a path toward more realistic PHY-layer wireless simulation by relying on waveform superposition and real demodulation. Building a practical, general-purpose emulator remains challenging, as it spans physical-layer signal processing, high-performance computing, hardware validation, and integration with evolving tools such as GNU Radio and Sionna RT. Our preliminary GPU-based results suggest that higher-bandwidth operation may be achievable, extending the scope beyond IoT-focused sub-GHz scenarios. Beyond direct technology evaluation, such a platform could support benchmarking under specific coexistence conditions, help prepare deployments by replaying realistic interference situations, and derive lower-resolution models from IQ-level signal mixing for analytical studies and larger-scale simulation. It could also support \emph{Reinforcement Learning (RL)} for learning transmission configurations through interaction with the emulated channel and receivers. These directions motivate continued development of IQSim as an open-source tool for the community, together with broader validation across waveform generation modes, resampling strategies, receiver implementations, real-world measurements, and overhead benchmarks against traditional simulators for comparable scenarios.

\bibliography{references}

\end{document}